\documentclass[pra,aps,twocolumn,superscriptaddress,showpacs]{revtex4}
\usepackage{graphicx}
\usepackage{amsmath}

\newcommand{\ket}[1]{|#1\rangle}
\newcommand{\bra}[1]{\langle#1|}

\newcommand{\eq}{\begin{equation}}
\newcommand{\fine}{\end{equation}}

\begin{document}

\title{Entanglement criteria for microscopic-macroscopic systems}
\author{Nicol\`{o} Spagnolo} 
\affiliation{Dipartimento di Fisica, Sapienza Universit\`{a} di Roma, piazzale Aldo Moro 5, I-00185 Roma, Italy}
\affiliation{Consorzio Nazionale Interuniversitario per le Scienze Fisiche della Materia, piazzale Aldo Moro 5, I-00185 Roma, Italy}
\author{Chiara Vitelli}
\affiliation{Dipartimento di Fisica, Sapienza Universit\`{a} di Roma, piazzale Aldo Moro 5, I-00185 Roma, Italy}
\affiliation{Consorzio Nazionale Interuniversitario per le Scienze Fisiche della Materia, piazzale Aldo Moro 5, I-00185 Roma, Italy}
\author{Fabio Sciarrino}
\email{fabio.sciarrino@uniroma1.it}
\homepage{http://quantumoptics.phys.uniroma1.it}
\affiliation{Dipartimento di Fisica, Sapienza Universit\`{a} di Roma, piazzale Aldo Moro 5, I-00185 Roma, Italy}
\affiliation{Istituto Nazionale di Ottica, Consiglio Nazionale delle Ricerche (INO-CNR), largo Fermi 6, I-50125 Firenze, Italy}
\author{Francesco De Martini} 
\affiliation{Dipartimento di Fisica, Sapienza Universit\`{a} di Roma, piazzale Aldo Moro 5, I-00185 Roma, Italy}
\affiliation{Accademia Nazionale dei Lincei, via della Lungara 10, I-00165 Roma, Italy}

\begin{abstract}
We discuss the conclusions that can be drawn on a recent experimental micro-macro entanglement test
[F. De Martini, F. Sciarrino, and C. Vitelli, Phys. Rev. Lett. \textbf{100}, 253601 (2008)]. The system under
investigation is generated through optical parametric amplification of one photon belonging to
an entangled pair. The adopted entanglement criterion makes it possible to infer the presence of entanglement
before losses, that occur on the macrostate, under a specific assumption. In particular, an \emph{a priori}
knowledge of the system that generates the micro-macro pair is necessary to exclude a class of
separable states that can reproduce the obtained experimental results.
Finally, we discuss the feasibility of a micro-macro ``genuine'' entanglement test on the analyzed
system by considering different strategies, which show that in principle a fraction
$\varepsilon$, proportional to the number of photons that survive the lossy process, of the original
entanglement persists in any losses regime.
\end{abstract}

\pacs{03.67.Mn, 03.65.Ud, 03.67.Bg}

\maketitle

\section{Introduction}

The observation of quantum phenomena, such as quantum
entanglement \cite{Horo09}, has been mainly limited to systems of only few
particles. One of the main open challenge for an experimental test
in systems of large size is the construction of suitable criteria
for the detection of entanglement in bipartite macroscopic
systems. Much effort has been devoted in the last few years in
this direction. Some of them, such as the partial transpose
criterion developed by Peres in Ref.\cite{Pere96}, require the
tomographic reconstruction of the density matrix, which for system
of a large number of particles becomes highly demanding from an
experimental point of view. In order to avoid the necessity of the
complete reconstruction of the state, a class of tests where only
few local measurements are performed has been introduced under the
name of ``entanglement witness'' \cite{Horo96}. For bipartite
systems of a large number of particles, this approach has been
further investigated considering the possibility to exploit
collective measurements on the multiparticle state. Within this
context, Duan et al. proposed a general criterion in
Ref.\cite{Duan00} based on continuous variable \cite{Brau05}
observables. This general criterion was subsequently applied to
the quantum extension of the Stokes parameters
\cite{Koro02,Schn03} to obtain an entanglement bound for such kind
of variables \cite{Koro05}. Other approaches have been developed
based on spin variables \cite{Simo03} or pseudo-Pauli operators
\cite{Chen02}. An experimental application of this criteria based
on collective spin measurements has been performed in a bipartite
system of two gas samples \cite{Juls01}. However, an experimental realization of most of
these criteria in the quantum optical domain requires
photon-number resolving detectors with unitary efficiency, which
is beyond the current technology. A feasible approach for the
analysis of multiphoton fields has been developed in the last few
years, and is based on the deliberate attenuation of the analyzed
system up to the single photon level. In this way, standard
single-photon techniques and criteria can be used to investigate
the properties of the field. The verification of the entanglement
in the high-loss regime is an evidence of the presence of
entanglement before the attenuation, since no entanglement can be
generated by local operations. Such approach has been exploited in
\cite{Eise04,Cami06} to demonstrate the presence of entanglement
in a high gain spontaneous parametric down-conversion source up to
12 photons. An analogous conclusion has been theoretically obtained
in Ref.\cite{Durk04} on the same system by exploiting symmetry
considerations of the source. The attenuation method has been also
applied to a different system, making it possible to obtain an experimental
proof of the presence of entanglement between a single photon
state and a multiphoton state generated through the process of
optical parametric amplification in an universal cloning
configuration up to 12 photons \cite{DeMa05}.

In this paper we discuss recent experimental results of a micro-macro
entanglement test \cite{DeMa08}, where the system under investigation is
realized through the process of optical parametric amplification \cite{DeMa98,DeMa05a}
of an entangled photon pair. The exploited entanglement criterion is an extension of the spin-based
single-particle criterion of Ref.\cite{Eise04}. Such an extension requires a
supplementary assumption which will be clarified in the remaining part of this paper.
In Sec.\ref{sec:micro_macro} we briefly review the properties of the micro-macro
system realized in Ref.\cite{DeMa08}. Then, in Sec.\ref{sec:entanglement_test}
we discuss in details the performed entanglement test. In particular,
we focus on the conditions adopted in order to justify the exploited
entanglement criterion.
Finally, in Sec.\ref{sec:general_criteria} we perform a theoretical analysis
of the micro-macro system based on the parametric amplification of an
entangled pair. Several approaches for the verification of the entanglement
property of the system will be addressed, showing that a substantial fraction $\varepsilon$
of the original entanglement survives even in high losses condition.

\section{Micro-Macro system by amplification of an entangled photon pair}
\label{sec:micro_macro}

\begin{figure*}[ht!]
\centering
\includegraphics[width=0.85\textwidth]{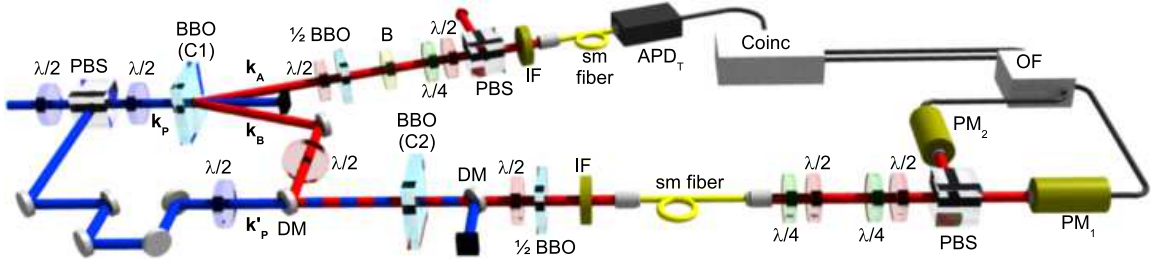}
\caption{(Color online) Scheme of the experimental setup. The main UV laser beam
provides the OPA excitation field beam at $\lambda_{P} = 397.5 nm$. A type II
BBO (Beta Barium Borate) crystal (crystal 1: C1) generates pair of photons with $\lambda = 795 nm$.
In virtue of the non-local correlations established between the modes
$\mathbf{k}_{A}$ and $\mathbf{k}_{B}$, the preparation of a single-photon on
mode $\mathbf{k}_{B}$ with polarization state $\vec{\pi}_{\varphi}$ is conditionally
determined by detecting a single-photon after proper polarization analysis on
the mode $\mathbf{k}_{A}$ [polarizing beamsplitter (PBS), $\lambda/2$ and
$\lambda/4$ waveplates, Soleil-Babinet compensator (B), interferential filter (IF), avalanche
photodiode (APD$_{T}$)].
The photon belonging to $\mathbf{k}_{B}$, together with the pump laser beam
$\mathbf{k}_{p}'$, is fed into an high gain optical parametric amplifier consisting
of a NL crystal 2 (C2), cut for collinear type-II phase matching. Measurement apparatus:
the field is analyzed by two photomultipliers (PM$_{1}$ and PM$_{2}$) and then discriminated through an
O-Filter device (OF), whose action is described in the text. For more details refer to \cite{DeMa08}.}
\label{fig:experimental_setup}
\end{figure*}

Let us first briefly review
the micro-macro system of Ref.\cite{DeMa08}. The system under investigation
is given by the following micro-macro source. An entangled pair of two photons
in the singlet state $\left\vert \psi ^{-}\right\rangle _{A,B}$=$2^{-{\frac{1}{2}}}
\left( \left\vert H\right\rangle _{A}\left\vert V\right\rangle _{B}-\left\vert V
\right\rangle_{A}\left\vert H\right\rangle _{B}\right)$ is produced through
spontaneous parametric down-conversion (SPDC) by crystal 1 (C1) pumped by a
pulsed UV pump beam: Fig.\ref{fig:experimental_setup}. There $\left\vert H\right\rangle $
and $\left\vert V\right\rangle $ stands, respectively, for a single photon
with horizontal and vertical polarization $(\vec{\pi}_{H,V})$ while
the labels $A,B$ refer to particles associated respectively with the spatial
modes $\mathbf{k}_{A}$and $\mathbf{k}_{B}$.
The photon belonging to $\mathbf{k}_{B}$, together with a strong UV pump beam, is injected
into an optical parametric amplifier (OPA) consisting of a non-linear crystal 2 (C2)
pumped by the beam $\mathbf{k}_{P}^{\prime}$. The crystal 2 is oriented for
collinear operation, i.e., emitting pairs of amplified photons over the same
spatial mode which supports two orthogonal $\vec{\pi}$ modes,
respectively horizontal and vertical. Then, fringe patterns are recorded by varying
the analyzed polarization in the single-photon site on the equatorial plane
of the Bloch sphere, and keeping fixed the analyzed polarization
in the $\mathbf{k}_{B}$ mode.
%
%
The interaction Hamiltonian of the OPA device is given by
$\hat{\mathcal{H}}_{OPA} = \imath \hbar \chi \hat{a}^{\dag}_{H}
\hat{a}^{\dag}_{V} + \mathrm{H.c.}$. In this collinear
configuration, the \emph{quantum injected} optical parametric
amplifier (QIOPA) acts as an \emph{optimal phase-covariant cloning
machine} \cite{Scia05,Scia07}, thus performing the optimal quantum
cloning process for all single photon states belonging to the
equatorial plane of the polarization Bloch sphere, defined as
$\vec{\pi}_{\phi} = 2^{-1/2} (\vec{\pi}_{H} + e^{\imath \phi}
\vec{\pi}_{V})$. For all equatorial basis, the interaction
Hamiltonian presents the same form: $\hat{\mathcal{H}}_{OPA} =
\imath \hbar \chi/2 e^{- \imath \phi} \left( \hat{a}^{\dag 2}_{\phi} - \hat{a}^{\dag
2}_{\phi_{\bot}} \right) + \mathrm{H.c.}$. The overall micro-macro
state after the amplification process reads:
\begin{equation}
\vert \Psi^{-} \rangle_{AB}  = \frac{1}{\sqrt{2}} (\vert \phi \rangle_{A} \vert \Phi^{\phi_{\bot}} \rangle_{B} -
\vert \phi_{\bot} \rangle_{A} \vert \Phi^{\phi} \rangle_{B})
\end{equation}
The output multiphoton states $\vert \Phi^{\phi} \rangle = \hat{U}_{OPA} \vert \phi \rangle$,
where $\vert \phi \rangle$ labels the injection of single-photon state with equatorial
polarization, is given by the following expression:
\begin{equation}
\vert \Phi^{\phi} \rangle = \frac{1}{C^{2}} \sum_{i,j=0}^{\infty} \gamma_{ij} \vert (2i+1)\phi, (2j)\phi_{\bot} \rangle
\end{equation}
where $\gamma_{ij} = (\frac{\Gamma}{2})^{i} (- \frac{\Gamma}{2})^{j} e^{- \imath (i+j) \phi} \frac{\sqrt{(2i+1)!(2j)!}}{i!j!}$,
$\Gamma = \tanh g$, $C = \cosh g$, with $g= \chi t$ nonlinear gain of the amplifier. Hereafter, the state
$\vert n\pi, m\pi_{\bot} \rangle$ labels a Fock state with $n$ photons with $\pi$ polarization and $m$ photons
with $\pi_{\bot}$ polarization.
For a detailed discussion on the properties of such states we refer to Refs.\cite{DeMa08,DeMa09}.

Symmetry considerations based on the \textit{rotational invariance} of
the overall micro-macro \textit{singlet} photon pair $\vert \psi^{-} \rangle$
and of the \textit{phase-covariant} and \textit{information preserving}
properties of the adopted QI-OPA,\ lead to
conclude that the two fringe pattern recorded in two different
equatorial basis, $\{+,-\}$ ($\phi=0$) and $\{ R,L \}$ ($\phi=\pi/2$) are identical,
in the sense that the micro and macro states adopted in both cases, present
the same Fock-space expansion. In practice, the experimental visibilities
of the fringe patterns in such two equatorial basis have been found equal
by \cite{DeMa08}, within the statistical errors.

\section{Entanglement test}
\label{sec:entanglement_test}
In this section we discuss a recent entanglement test performed in Ref.\cite{DeMa08}. The system under
investigation is the micro-macro source discussed in the previous section. We focus our analysis on the
exploited entanglement criterion, obtained as the extension of a spin-based criterion for a bipartite
microscopic-microscopic system \cite{Eise04}. First, the criterion and its experimental
implementation is introduced in details, including an analysis of the regions of the system's Hilbert space filtered by the detection strategy and a numerical analysis of the effects of a lossy process. Then, the assumptions on the source necessary for the validity of the test are discussed.

\subsection{Micro-micro entanglement witness}
For a two-photon state $\ket{\psi}$, defined on two different modes $a$ and $b$, the entanglement is demonstrated by applying the following criterion. For any
separable state, the following inequality holds \cite{Durk04a,Eise04}:
\begin{equation}\label{eq:micro_micro_criterion}
_{\psi}\langle\hat{\sigma}_{1}^{(a)}\otimes\hat{\sigma}_{1}^{(b)} \rangle_{\psi}+_{\psi} \langle
\hat{\sigma}_{2}^{(a)}\otimes\hat{\sigma}_{2}^{(b)} \rangle_{\psi}+_{\psi} \langle \hat{\sigma}_ {3}^{(a)}\otimes\hat{\sigma}_{3}^{(b)} \rangle_{\psi} \leq 1
\end{equation}
\noindent where $\hat{\sigma}_{1,2,3}$ are the Pauli operators and $_{\psi} \langle \cdot
\rangle_{\psi}$ stands for the average on the state $\ket{\psi}$.

\subsection{Micro-macro entanglement witness in the ideal case}
The same criterion can be extended to a micro-macro scenario, by
measuring the pseudo spin operators $\hat{\Sigma}_{i}$ on the
macro state, obtained through an unitary transformation upon the
micro-micro state. Here, the $\hat{\Sigma}_{i}$ operators are the
time evolution of the Pauli operators according to
$\hat{\Sigma}_{i} = \hat{U} \hat{\sigma}_{i} \hat{U}^{\dag}$,
where $\hat{U}$ is the time evolution operator of the amplifier $\hat{U}_{OPA}$.
The following inequality holds:
\begin{equation}
\label{eq:micro_macro_criterion}
_{\Psi}\langle\hat{\sigma}_{1}^{(a)}\otimes\hat{\Sigma}_{1}^{(b)} \rangle_{\Psi}+_{\Psi} \langle
\hat{\sigma}_{2}^{(a)}\otimes\hat{\Sigma}_{2}^{(b)} \rangle_{\Psi}+_{\Psi} \langle \hat{\sigma}_ {3}^{(a)}\otimes\hat{\Sigma}_{3}^{(b)} \rangle_{\Psi} \leq 1
\end{equation}
\noindent where $i=1,2,3$ refer to the polarization basis $1\rightarrow \{H,V\}$, $2\rightarrow \{R,L\}$, $3\rightarrow \{+,-\}$.
Since the operators $\hat{\Sigma}_{i}$ are built from the unitary evolution of eigenstates
of $\hat{\sigma}_{i}$, they satisfy the same commutation rules of the single-particle 1/2-spin: $\left[ \hat{\Sigma}_{i}, \hat{\Sigma}_{j} \right] = �
2 \imath \epsilon_{ijk} \hat{\Sigma}_{k}$, where $\epsilon_{ijk}$ is the Levi-Civita tensor density. Indeed we have for $i=1$:
\begin{equation}
\label{eq:pauli_1}
\begin{aligned}
\hat{\Sigma}_{1} &= \sum_{n,m=0}^{\infty} \gamma_{n} \gamma_{m}^{\ast} \vert (n+1)H,nV \rangle \langle (m+1)H,mV \vert +\\
&- \sum_{n=0}^{\infty} \gamma_{n} \gamma_{m}^{\ast} \vert nH,(n+1)V \rangle \langle mH,(m+1)V \vert
\end{aligned}
\end{equation}
where $\gamma_{n} = \frac{\Gamma^{n}}{C^{2}} \sqrt{n+1}$. For $i=2,3$ we have:
{\small
\begin{equation}
\label{eq:pauli_23}
\begin{aligned}
\hat{\Sigma}_{i} &= \sum_{n,m,p,q=0}^{\infty} \gamma^+_{nm} \gamma_{pq}^{+ \ast}\vert (2n+1)\vec{\pi}_{i}, (2m)\vec{\pi}^{\bot}_{i} \rangle \langle (2p+1)\vec{\pi}_{i}, (2q)\vec{\pi}^{\bot}_{i} \vert\\
&- \sum_{n,m,p,q=0}^{\infty} \gamma^-_{nm} \gamma_{pq}^{- \ast}  \vert (2n)\vec{\pi}_{i}, (2m+1)\vec{\pi}^{\bot}_{i} \rangle \langle (2p)\vec{\pi}_{i}, (2q+1)\vec{\pi}^{\bot}_{i} \vert
\end{aligned}
\end{equation}
}where $\gamma_{nm}^+= \frac{1}{C^{2}} (\frac{\Gamma}{2})^{n} (-\frac{\Gamma}{2})^{m} \frac{\sqrt{(2n+1)!(2m)!}}{n!m!}$ and $\gamma_{nm}^-= \frac{1}{C^{2}} (\frac{\Gamma}{2})^{n}
(-\frac{\Gamma}{2})^{m} \frac{\sqrt{(2n)!(2m+1)!}}{n!m!}$.

\noindent In equation (\ref{eq:micro_macro_criterion}) the state
$\ket{\Psi}$ is obtained by the amplification of the state
$\ket{\psi}$ over the single spatial mode $\mathbf{k_{B}}$, and can be identified as two-qubit state of micro and macro
systems. In the ideal
case, the following map holds:
\begin{eqnarray}
\ket{\pm} & \rightarrow & \ket{\Phi^{\pm}}=U\ket{\pm}\nonumber\\
\ket{R/L} & \rightarrow & \ket{\Phi^{R,L}}=U\ket{R/L}
\end{eqnarray}
\noindent where $\hat{U}$ is the unitary amplification operator.

\subsection{Micro-macro entanglement in the lossy case}

\begin{figure*}[ht!]
\includegraphics[width=0.8\textwidth]{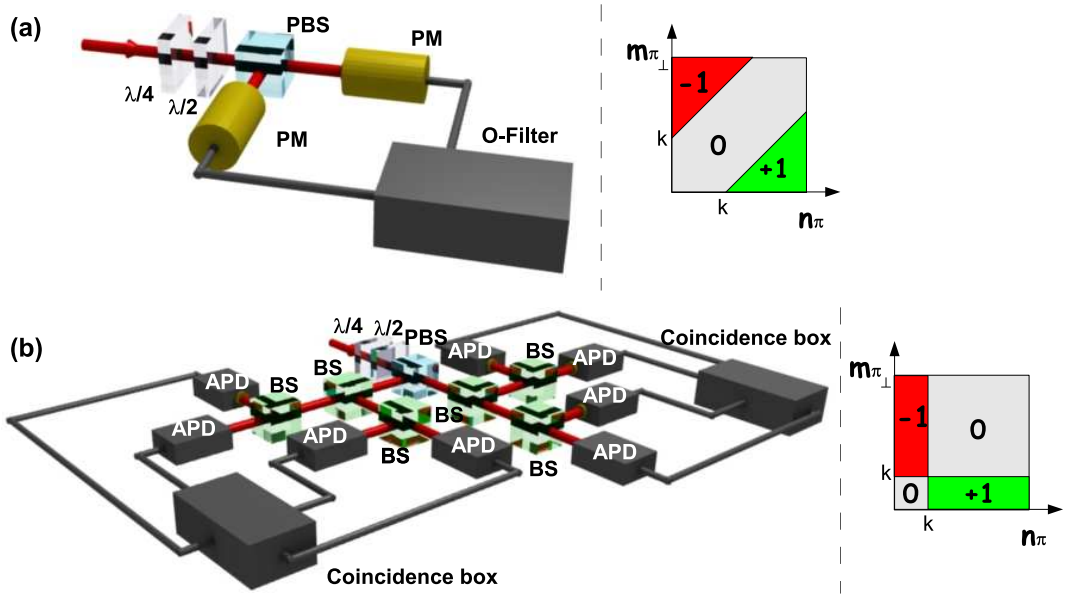}
\caption{(Color online) (a) O-Filter based detection apparatus. The field is analyzed in polarization [$\lambda/4$ and 
$\lambda/2$ wave-plates, polarizing beam-splitter (PBS)] and the intensities are measured by two 
photomultipliers (PM).  Right figure: diagram of the two-mode Fock space's region selected by the O-Filter
measurement scheme. Green region (+1) corresponds to the condition $n_{\pi} - m_{\pi_{\bot}} > k$, red region (-1) 
corresponds to the condition $m_{\pi_{\bot}} - n_{\pi} > k$, grey region (0) corresponds to the condition 
$\vert n_{\pi} - m_{\pi_{\bot}} \vert < k$. (b) Multi-detector measurement strategies. The field should be analyzed 
in polarization [$\lambda/4$ and $\lambda/2$ wave-plates, polarizing beam-splitter (PBS)]. Each polarization 
state should be divided in equal parts by a sequence of $50/50$ beam-splitters (BS) and the detected by four APD's 
(Avalanche photo-diodes): the coincidences between all four detectors trigger the successful events.
Right figure: diagram of the two-mode Fock space's region selected by the multi-detector measurement scheme. 
Green region (+1) corresponds to the presence of a coincidence only between all $\pi$ polarization detectors, red region 
(-1) corresponds to presence of a coincidence only between all $\pi_{\bot}$ polarization detectors, grey region 
(0) corresponds to the inconclusive outcome. In this case, k is the number of detectors.}
\label{fig:Ofilter}
\end{figure*}

\textsl{(a) Implementation of the pseudo-Pauli operators: the O-Filter.} 
Since measurements of  Eqs.(\ref{eq:pauli_1}-\ref{eq:pauli_23}), which require the perfect discrimination 
of the number of photons present in the detected state,  are out of reach by current
technology, we have adopted another strategy which is based on the
 O-Filter (OF) device, shown in Fig.\ref{fig:Ofilter}-(a). 
This method is based on a probabilistic discrimination 
of the macro-states $\vert \Phi^{\phi} \rangle$ and $\vert \Phi^{\phi_{\bot}} 
\rangle$, which exploits the macroscopic features present in their photon-number 
distributions.

Such measurement is implemented by an intensity measurement in the
$\{ \vec{\pi}_{i}, \vec{\pi}^{\bot}_{i} \}$ basis, followed by an
electronic processing of the signal. If
$n_{\pi}-m_{\pi_{\bot}}>k$, the (+1) outcome is assigned to the
event, if $m_{\pi_{\bot}}-n_{\pi}>k$ the (-1) outcome is assigned
to the event. If $\vert n_{\pi}-m_{\pi_{\bot}} \vert <k$, an
inconclusive outcome (0) is assigned to the event. The action of
the O-Filter is described by the following measurement
observables, applied on the multiphoton state \textsl{after}
losses:
\begin{eqnarray}
\label{eq:O-Filtering_POVM_1}
\hat{\Pi}_{i}(k) &=&\sum_{n=k}^{\infty }\sum_{m=0}^{n-k}
\vert n \vec{\pi}_{i}, m \vec{\pi}^{\bot}_{i} \rangle \langle n \vec{\pi}_{i}, m\vec{\pi}^{\bot}_{i} \vert -\\
\nonumber &\;&\sum_{m=k}^{\infty }\sum_{n=0}^{m-k}%
\vert n \vec{\pi}_{i}, m \vec{\pi}^{\bot}_{i} \rangle \langle n \vec{\pi}_{i}, m\vec{\pi}^{\bot}_{i} \vert
\end{eqnarray}
The state after losses is no more a macro-qubit living in
a two dimensional Hilbert space, but in general it is represented by
the density matrix $\hat{\rho}_{\eta}^{\phi}$. Such density matrix
is obtained by applying to the macroqubit $\vert \Phi^{\phi} \rangle$
the map that describes the action a lossy channel with transmittivity 
$\eta$: $\mathcal{L}[\hat{\rho}] = \sum_{p} \gamma_{p} \hat{a}^{p} 
\hat{\rho} \hat{a}^{\dag \, p} \gamma^{\dag}_{p}$ where $\gamma_{p} = 
\frac{1}{\sqrt{p!}} (1-\eta)^{p/2} \eta^{(\hat{a}^{\dag} \hat{a})/2}$ 
\cite{Durk04a}.  In order to describe the measurement results, all
the average values of the measurement operators must be calculated
with the density matrix of the state after losses $\hat{\rho}_{\eta}^{\phi}$.
Nevertheless the large difference in the photon number distribution 
along the distribution's tails present in the macro-qubits before losses
is present also in the distribution of the macro-states after losses.
A detailed discussion on the properties of the macrostates after losses
in both the Fock-space and the phase-space is reported in Refs.
\cite{DeMa09,Spag09}. By exploiting this feature of our system, this
probabilistic detection method allows us to infer the generation
\textsl{before} losses of a $\vert \Phi^{\phi} \rangle$ or a
$\vert \Phi^{\phi_{\bot}} \rangle$ state by exploiting the
information encoded in the unbalancement of the number of photons
present in the state \textsl{after} losses $\hat{\rho}_{\eta}^{\phi}$.

An analogous measurement scheme is shown in Fig.\ref{fig:Ofilter}-(b). 
The field is analyzed in polarization, and each branch is equally 
divided among a set of single-photon detectors (APD). Coincidences 
between the output TTL signals are recorded for each analyzed polarization, 
and the (+1) or the (-1) outcomes are assigned depending on which of the 
two analyzed sets of APDs record the $N$-fold coincidence. If no $N$-fold 
coincidences are recorded, the (0) inconclusive outcome is assigned to 
the event. This scheme performs the measurement of the $N$-th order 
correlation function of the field, where $N$ is the number of detectors. 
We note that the O-Filter based and the multi-detector based schemes select 
analogous regions of the Fock space.


\begin{figure}[b!]
\includegraphics[width=0.5\textwidth]{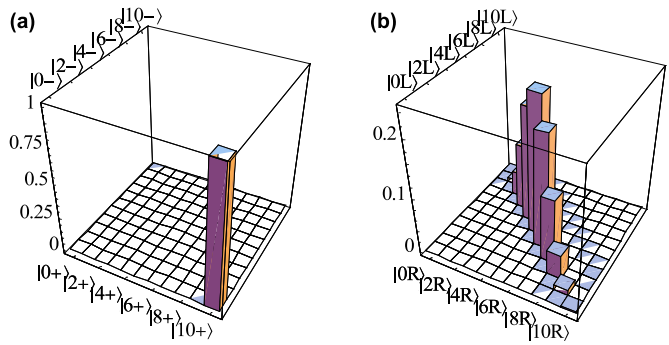}
\caption{(Color online) (a) Photon number distribution for a Fock state $\vert
n+, 0- \rangle$ with $n=10$ in the $\{+,-\}$ polarization basis.
(b) Photon number distribution for a Fock state $\vert n+, 0-
\rangle$ with $n=10$ in the $\{R,L\}$ polarization basis.}
\label{fig:filtering}
\end{figure}

\textsl{(b) Filtering of the detected state.} The entanglement test
performed on our system in Ref.\cite{DeMa08} is given by 
Eq.(\ref{eq:micro_macro_criterion}) where the $\hat{\Sigma}$
operators are replaced with the $\hat{\Pi}$ operators of the
O-Filter:
\begin{equation}
\label{eq:micro_macro_criterion_OF}
_{\Psi}\langle\hat{\sigma}_{1}^{(a)}\otimes\hat{\Pi}_{1}^{(b)} \rangle_{\Psi}+_{\Psi} 
\langle \hat{\sigma}_{2}^{(a)}\otimes\hat{\Pi}_{2}^{(b)} \rangle_{\Psi}+_{\Psi} 
\langle \hat{\sigma}_{3}^{(a)}\otimes\hat{\Pi}_{3}^{(b)} \rangle_{\Psi} \leq 1
\end{equation}
\noindent It is worth noting that, in general, the resulting Eq.(\ref{eq:micro_macro_criterion_OF}) 
is no longer an entanglement witness. As discussed in the remaining of the paper, the bound of Eq.(\ref{eq:micro_macro_criterion_OF}) 
can be recovered as an entanglement witness by making a supplementary assumpion on the micro-macro source.
On one side, we note that the measurement of the correlations for the entanglement test
of Eq.(\ref{eq:micro_macro_criterion_OF}), are performed in the same basis for Alice and Bob's
sites. However, care should be taken when a filtering of the
detected state is performed. As shown in
Fig.\ref{fig:Ofilter}-(a), the O-filter detection scheme
corresponds to a Fock space filtering of the output state. The
measurements performed on different polarization basis select
\textsl{different} regions of the Fock space, corresponding to
different portions of the density matrix. This is shown in
Fig.\ref{fig:filtering}, where the photon number distribution of a
$\vert n+, 0- \rangle$ Fock state with $n=10$ in the $\{+,-\}$ and
$\{R,L\}$ polarization bases is reported. When measured with the
O-Filter device, such state generates a conclusive (+1) outcome in
the $\{+,-\}$ basis, since a strong unbalancement is present
between the two polarizations. On the contrary, in the $\{R,L\}$
basis with high probability the state generates an inconclusive
outcome (0) and is filtered out. This feature has no counterpart
in the micro-qubit formalism: indeed the Hilbert space of the
original photon is only two-dimensional, so there is no risk of
different subspaces being detected for different choices of
measurement basis. Indeed the presence of losses enlarges
the dimension of the Hilbert space in which the macro-qubit lives,
and the criterion of Eq.(\ref{eq:micro_macro_criterion_OF}) requires 
an auxiliary assumptions on the micro-macro state.

\textsl{(c) Violation of the entanglement bound for a micro-macro
separable state.} Without any assumption on the investigated system 
the inequality (\ref{eq:micro_macro_criterion_OF}), that is, the 
original pseudo-Pauli criterion (\ref{eq:micro_macro_criterion}) where
the $\{\hat{\Sigma}_{i} \}$ operators have been replaced by the 
$\{\hat{\Pi}_{i} \}$ ones, does not represent anymore a bound for
entangled states. It is satisfied by separable states of the form
\cite{Seka09}:
\begin{eqnarray}\label{eq:separable}
\hat{\rho}_{sep} &=&\frac{1}{2\pi}\int_{0}^{2\pi}d\phi
\hat{U}(\phi)\ket{1\pi_{i},0\pi_{i}^{\perp}}_{a}\ket{0\pi_{i},N\pi_{i}^{\perp}}_{b}\times\nonumber\\
&\;& _{a}\bra{1\pi_{i},0\pi_{i}^{\perp}}_{b}\bra{0\pi_{i},N\pi_{i}^{\perp}}\hat{U}(\phi)^{\dag}
\end{eqnarray}
\noindent where $\hat{U}(\phi)$ is a rotation of the whole system polarization around the $z$ axis by an angle $\phi$.\\

\subsection{Auxiliary assumption on the micro-macro system}
Despite the previous considerations, the OF based strategy allows us to discriminate between different
macro states in a probabilistic way.
\begin{figure}[ht!]
\includegraphics[width=0.4\textwidth]{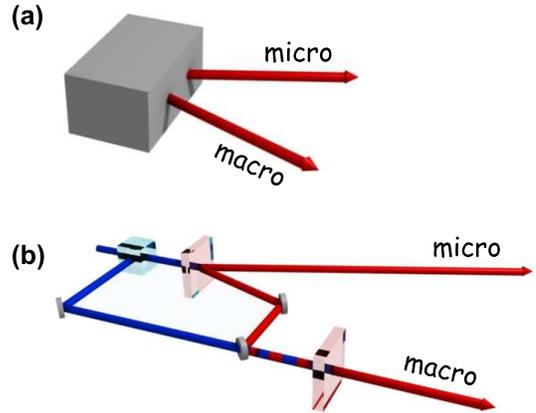}
\caption{(Color online) (a) Micro-macro system source in a black box configuration: no assumption is made about the source.
(b) Micro-macro amplified system: the macroscopic state is generated by a coherent amplification process of a single
photon, belonging to an entangled pair in the singlet polarization state $\vert \psi^{-} \rangle$.}
\label{fig:figure_assumption}
\end{figure}
When $k\rightarrow \infty$ the mean value of the $\hat{\Pi}_{i}$
operators, calculated over the real state
$\mathrm{Tr}(\hat{\rho}^{\phi}_{\eta}\hat{\Pi}_{i})$, tends to the
mean value of the Pauli pseudo-spin operators, calculated over the
ideal macro-qubit one
$\bra{\Phi^{\phi}}\hat{\Sigma}_{i}\ket{\Phi^{\phi}}$. Indeed, for
asymptotically high values of the threshold $k \rightarrow
\infty$, the measurement of the $\hat{\Pi}_{i}$ operators on the
$\hat{\rho}_{\eta}^{\phi}$ allows perfect, although probabilistic
in the spirit of positive operator-valued measurements (POVM), 
discrimination of orthogonal states, as the
pseudo-spin operator does for the macro-qubits $\vert \Phi^{\phi}
\rangle$. In other words, if the $\vert \Phi^{+} \rangle$ state is
generated, the measurement with the $\hat{\Sigma}_{i}$ operator in
the $\{ +,- \}$ basis never leads to the (-1) outcome. At the same
time, the measurement of the $\hat{\rho}_{\eta}^{+}$ state after
losses with the $\hat{\Pi}_{i}$ operator in the $\{ +,- \}$ basis
does not generate the (-1) outcome if $k$ is large enough.
According to these considerations, we can infer the presence of
the macro-qubit \textsl{before} losses and \textsl{after} the
amplifier and then apply the original micro-macro inequality of
Eq.(\ref{eq:micro_macro_criterion}). This inference implies an
assumption on the micro-macro system: the macro state has to be
generated by an amplification process upon a  micro-micro
entangled  pair. The difference between the general case
of a micro-macro entangled setup and the one here described is
pointed out in fig.\ref{fig:figure_assumption}-(a) and
fig.\ref{fig:figure_assumption}-(b) respectively. Therefore the
entanglement test performed by the OF scheme allows us to infer
the presence of entanglement \textsl{at least before} losses, and
to demonstrate the capability of amplifying an entangled pair in a
coherent way. Indeed the class of separable state in
Eq.(\ref{eq:separable}) cannot be generated by a coherent
amplification process, and the coherence of amplification
is furthermore demonstrated by the presence of interference fringes
in two different polarization bases.

\subsection{Properties of the O-Filter detection strategy}
\begin{figure}[ht!]
\centering
\includegraphics[width=0.35\textwidth]{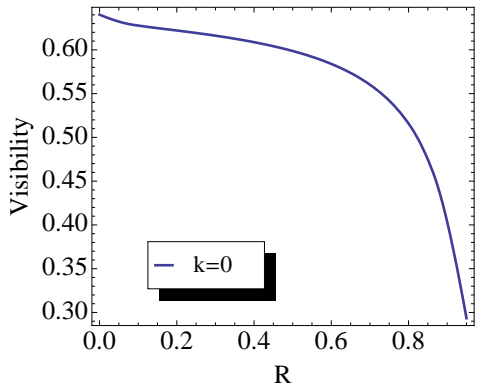}
\caption{(Color online) Trend of the visibility as a function of the losses parameter $R$ for a threshold $k=0$; $\langle n \rangle \sim 35$.}
\label{fig:visibility_Ofilter_a}
\end{figure}

\begin{figure}[ht!]
\centering
\includegraphics[width=0.35\textwidth]{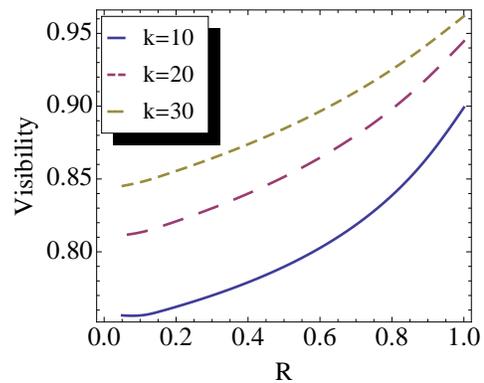}
\caption{(Color online) Trend of visibility as a function of the losses parameter $R$ for $k=10$ (solid line), $k=20$ (dashed line)
and $k=30$ (short-dashed line). All lines correspond to $\langle n \rangle \sim 35$.}
\label{fig:visibility_Ofilter_b}
\end{figure}

In order to conclude our analysis on the O-filtering measurement technique, we calculate theoretically
how the visibility of the fringe pattern obtained in the micro-macro amplified scheme scales with the 
amount of losses when it is measured with this detection strategy. The visibility is defined as:
\begin{equation}
V = \frac{P(+1) - P(-1)}{P(+1) + P(-1)}
\end{equation}
where $P(+1)$ and $P(-1)$ are the probability of obtaining respectively the (+1) and the (-1) outcome, and are calculated as:
{\small
\begin{eqnarray}
P(+1) &=& \mathrm{Tr} \left[ \hat{\rho}_{\eta}^{\phi} \left(\sum_{n=k}^{\infty }\sum_{m=0}^{n-k}
\vert n \vec{\pi}_{i}, m \vec{\pi}^{\bot}_{i} \rangle \langle n \vec{\pi}_{i}, m\vec{\pi}^{\bot}_{i} \vert
\right) \right]  \\
P(-1) &=& \mathrm{Tr} \left[ \hat{\rho}_{\eta}^{\phi} \left(\sum_{m=k}^{\infty }\sum_{n=0}^{m-k}
\vert n \vec{\pi}_{i}, m \vec{\pi}^{\bot}_{i} \rangle \langle n \vec{\pi}_{i}, m\vec{\pi}^{\bot}_{i} \vert
\right) \right]
\end{eqnarray}
}We consider separately two different cases.
(1) For a threshold $k=0$, no filtering is performed on the analyzed state since the complete
Fock space is selected by the O-filter device. In this case, the visibility is a \textsl{decreasing}
function of the losses parameter $R = 1 - \eta$, where $\eta$ is the overall quantum efficiency of
the channel. In Fig.\ref{fig:visibility_Ofilter_a} the trend of the visibility as a function of $R$ for $k=0$ and
a gain value $g=1.8$ is reported. We note that the visibility \textsl{decreases} with $R$ since
a larger amount of losses is responsible for the cancellation of a larger amount of entanglement.
This point will be clarified later in this paper.
(2) For a threshold $k>0$, the O-filter device performs a filtering
of the detected state as discussed in the previous paragraphs. In this case, the visibility is an
\textsl{increasing} function of the losses parameter $R = 1 - \eta$. In Fig.\ref{fig:visibility_Ofilter_b}
the trend of the visibility as a function of $R$ for a gain value of $g=1.8$, corresponding to an average
number of generated photons $\langle n \rangle \sim 35$, and several values of the threshold $k$.
The visibility \textsl{increases} with the losses parameter since, for the same value of the threshold $k$,
a tighter filtering of the detected wave function is performed for higher $R$ due to the reduced average
number of photons present in the state.\\

\section{General criteria for micro-macro entanglement}
\label{sec:general_criteria}
In this section we analyze different approaches for an entanglement test in a microscopic-macroscopic
system, and we discuss the application to the system of Ref.\cite{DeMa08} obtained by optical
parametric amplification of an entangled photon pair. As a first step, we develop a generalized
entanglement witness criterion based on dichotomic measurements. We then apply this criterion for
the specific case of the Pauli pseudo-spin operators previously introduced, showing the fragility
under losses of this detection strategy. As a second step, we consider a different approach based
on the deliberate attenuation of the multiphoton field. Such technique allows us to theoretically
demonstrate the presence of entanglement in the investigated micro-macro system for any value of
the amplifier gain and of the losses $\eta$. The same conclusion can be drawn with a different approach
based on the quantum Stokes operators already developed in \cite{Simo03} and applied to our micro-macro
system in Refs.\cite{Seka09,Seka10}.

\subsection{Generalized entanglement witness}
For a set $\{ \hat{D}_{i}\}$ of dichotomic operators, without making any supplementary assumption, the bound to be violated in order to demonstrate the entanglement of the overall micro-macro system must be modified with respect to Eq.(2), and a necessary condition for separable states is given by the following inequality:
\begin{equation}
\label{eq:general_criterion}
S = \langle \hat{\sigma}_{1}^{(a)} \otimes \hat{D}_{1}^{(b)} \rangle_{\Psi} + \langle
\hat{\sigma}_{2}^{(a)} \otimes \hat{D}_{2}^{(b)} \rangle_{\Psi} + \langle
\hat{\sigma}_{3}^{(a)} \otimes \hat{D}_{3}^{(b)} \rangle_{\Psi} \leq \sqrt{3}
\end{equation}
Details over the derivation of this criterion are reported in App.\ref{app:general_criterion}.
Such criterion presents the interesting feature of not requiring any knowledge of the Hilbert space
where the analyzed states live. Indeed, in the derivation of the bound (\ref{eq:general_criterion})
the only necessary assumption concerns the measurement operators, which can have only
two possible outcomes $(\pm 1)$.
\begin{figure}[ht!]
\centering
\includegraphics[width=0.4\textwidth]{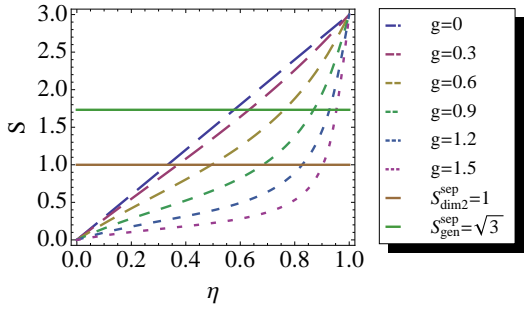}
\caption{(Color online) Numerical evaluation of the witness $S$ for the specific choice of the Pauli pseudo-spin
operators $\{ \hat{\Sigma}_{i}\}$ as measurement operators $\{ \hat{D}_{i} \}$ as a function of the detection
losses $\eta$, calculated for several values of the gain of the amplifier. The upper horizontal solid line
corresponds to the bound for separables states of the general criterion (\ref{eq:general_criterion}), while
the lower horizontal solid line corresponds to the bound for separable states (\ref{eq:micro_macro_criterion})
where the commutation properties of the operators have been exploited and a standard assumption on 
the Hilbert space is necessary. Dashed lines: from top to bottom, curves correspond respectively to a nonlinear gain
$g=0$, $g=0.3$, $g=0.6$, $g=0.9$, $g=1.2$ and $g=1.5$}
\label{fig:Witness_Pauli_micro_macro_p=1}
\end{figure}
We then applied the obtained criterion to evaluate the quantity $S$ for the micro-macro state generated
through the process of optical parametric amplification, for the specific choice of the Pauli pseudo-spin
operators $\{ \hat{\Sigma}_{i} \}$ (\ref{eq:pauli_1}-\ref{eq:pauli_23}) as the measurement operators.
More specifically, we evaluated the value of $S$ as a function of the transmission efficiency $\eta$
of the multiphoton mode $\mathbf{k}_{B}$ for several values of the gain $g$
[Fig.\ref{fig:Witness_Pauli_micro_macro_p=1}]. The value of $S$ is then compared to the bound for
separable states $S_{gen}^{sep}=\sqrt{3}$. We observe that this entanglement measurement is
fragile under losses, since the value of $S$ falls below the bound for separable
states when the number of lost photons is $R \langle n \rangle \sim 1$. Such result is expected
since the Pauli operators allows to distinguish the $\vert \Phi^{\phi} \rangle$ states exploiting the
well-defined parity in the number of photon generated by the amplifier depending on the polarization
of the input states. In presence of losses, such well-defined parity is quickly cancelled, thus not allowing
to discriminate among the macro-states with this kind of measurement. This feature of the macro-states
generated through the process of optical parametric amplification is reported and discussed in Refs.\cite{DeMa09a,Spag09}.

\subsection{Entanglement detection in a highly attenuated scenario}
\begin{figure}[ht!]
\includegraphics[width=0.5\textwidth]{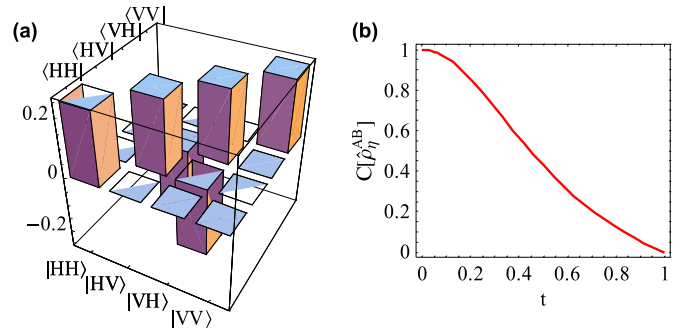}
\caption{(a) Density matrix of joint micro-macro system in the high losses regime, for a gain value of $g=3$ and a value of the losses parameter $\eta = 10^{-4}$. (b) Plot of the concurrence $C(\hat{\rho}_{\eta}^{AB})$ as a function of the parameter $t= \Gamma (1-\eta)$. We note the persistance of the off-diagonal terms and entanglement for all values of $g$ and $\eta$.}
\label{fig:concurrence_p1}
\end{figure}

An alternative approach can be used to demonstrate the presence of entanglement in our micro-macro configuration. The macroscopic field is deliberately attenuated up to the single-photon regime and detected through an APD. Such method has been exploited to demonstrate the entanglement up to 12 photons in a spontaneous parametric down conversion source \cite{Eise04}, or in a micro-macro configuration \cite{DeMa05}.
The average number of photons impinging onto the detector in this regime is then $\eta \langle n \rangle \leq 1$, where $\eta$ is the overall quantum efficiency of the channel. In this condition, the probability of detecting more then one photon becomes negligible. The density matrix of the macroscopic state can be reduced to a 1-photon
subspace, and the joint micro-macro system is defined in a $2 \times 2$ polarization Hilbert space spanned by the basis vectors $\{ \vert H \rangle_{A} \vert H \rangle_{B}, \vert H \rangle_{A} \vert V \rangle_{B}, \vert V \rangle_{A}, \vert H \rangle_{B}, \vert V \rangle_{A} \vert V \rangle_{B} \}$. The complete state $\hat{\rho}^{AB}_{\eta}$ can be then evaluated by applying the map describing a lossy channel \cite{Durk04} to the micro-macro amplified state $\hat{\rho}^{AB}_{\eta} =(\hat{I}^{A} \otimes \mathcal{L}_{\eta}^{B})\left[ (\hat{I}^{A} \otimes \hat{U}^{B}_{OPA}) \vert \psi^{-} \rangle_{AB} \langle \psi^{-} \vert (\hat{I}^{A} \otimes \hat{U}^{B}_{OPA})\right]$. We obtain the following expression:
\begin{equation}
\hat{\rho}_{\eta}^{AB}=\frac{1}{1+3t^{2}}\left(
\begin{array}{cccc}
t^{2} & 0 & 0 & 0 \\
0 & \frac{1}{2}\left( 1+t^{2}\right) & -\frac{1}{2}\left( 1+t^{2}\right) & 0
\\
0 & -\frac{1}{2}\left( 1+t^{2}\right) & \frac{1}{2}\left( 1+t^{2}\right) & 0
\\
0 & 0 & 0 & t^{2}
\end{array}
\right)
\end{equation}
where:
\begin{equation}
t=(1-\eta) \Gamma
\end{equation}
In Fig.\ref{fig:concurrence_p1}-(a) we show the density matrix of the joint micro-macro system for a value of $g=3.$ and $\eta = 10^{-4}$, showing the presence of the off-diagonal terms even in the high losses regime.
This system is entangled for any value of the nonlinear gain $g$. This property can be tested by application of the Peres criterion or by direct calculation of the concurrence, which reads:
\begin{equation}
\label{eq:conc}
C(\hat{\rho}_{\eta}^{AB})=\left( \frac{1-t^{2}}{1+3 t^{2}}\right) >0
\end{equation}

\begin{figure}[ht!]
\centering
\includegraphics[width=0.5\textwidth]{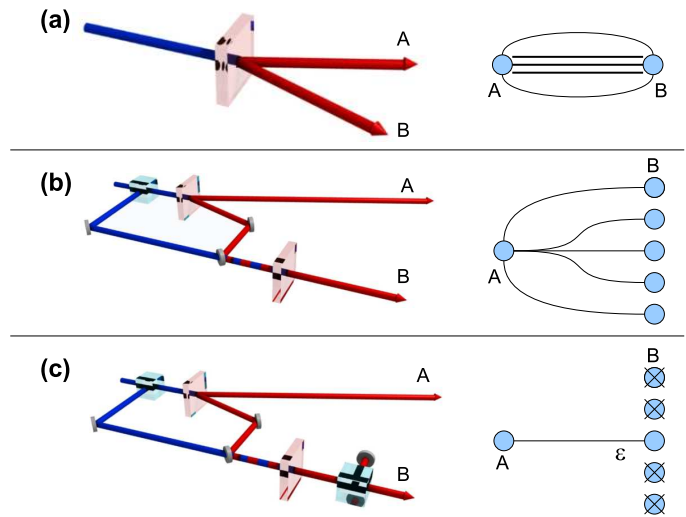}
\caption{(Color online) Diagramatic scheme of the entanglement broadcasting from the single photon pair to the multiparticle state. In presence of losses, the entanglement is reduced of a factor $\varepsilon$.}
\label{fig:entanglement}
\end{figure}

\begin{figure}[ht!]
\includegraphics[width=0.5\textwidth]{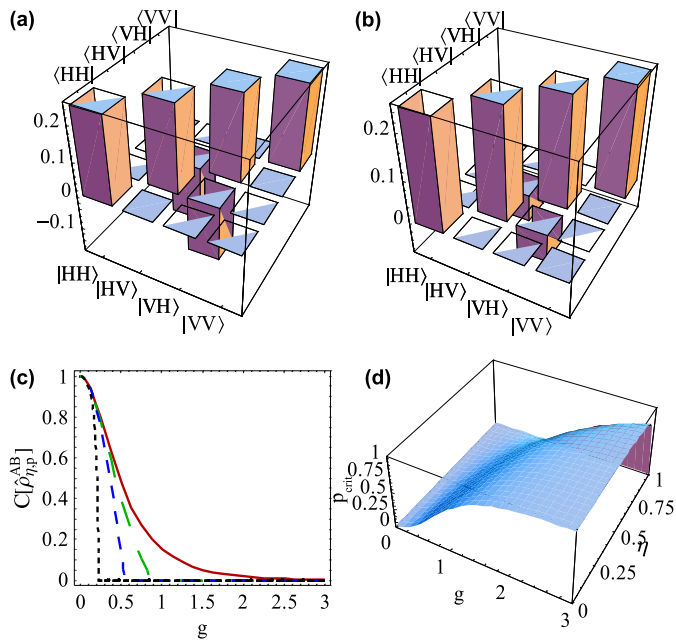}
\caption{(Color online) (a)-(b) Density matrix of the micro-macro system in the high losses regime, for a gain value of $g=3$ and a value of the losses parameter $\eta = 10^{-4}$. (a) Injection probability
$p=0.5$ and (b) injection probability of $p=0.15$. (c) Concurrence $C(\hat{\rho}_{\eta,p}^{AB})$ as a function of the gain $g$ for $\eta=10^{-4}$. Red solid line corresponds to an injection probability $p=1$, green long dashed line to $p=0.5$, blue short dashed line to $p=0.25$ and black dotted line to $p=0.05$. (d) 3-dimensional plot for the critical injection probability $p_{crit}$ as a function of the gain $g$ and the transmission coefficient $\eta$.}
\label{fig:concurrence_pmin1}
\end{figure}

This quantity is always positive, as plotted in Fig.\ref{fig:concurrence_p1}-(b), showing the presence of entanglement for any value of the gain. Since no entanglement can be generated with local operations (such as a lossy process) \cite{Eise04}, the presence of entanglement in the highly attenuated regime is due to the presence of entanglement in the micro-macro system before losses. \\
\noindent This criterion allows us to discuss an important feature of the micro-macro system based on optical parametric amplification. The entanglement of this system is generated in the micro-micro
source, where the singlet polarization state $\vert \psi^{-} \rangle$ is produced. The action of the amplifier is to broadcast the properties of the injected seed to the multiparticle state. In particular, the entanglement present in the original photon pair after the amplification process is transfered and shared among the generated particles (see Fig.\ref{fig:entanglement}). If a certain amount of losses is introduced in the macro-state and $\varepsilon$ is the percentage of photons that survive such decoherence process, the amount of entanglement detected after losses is reduced of a factor $\varepsilon$ but drops to 0 only if all particles are lost. Analytically, this feature is obtained by analyzing the expression (\ref{eq:conc}) for $C(\hat{\rho}_{\eta}^{AB})$. In the high gain limit ($\Gamma \sim 1$), the concurrence of our system in the highly attenuated regime becomes: 
\begin{equation}
C(\hat{\rho}_{\eta}^{AB}) \sim \frac{1-\Gamma^{2}}{1 + 3\Gamma^{2}} + \eta \frac{8 \Gamma^{2}}{(1 + 3 \Gamma^{2})^{2}} \stackrel{\Gamma \rightarrow 1}{\rightarrow} \frac{\eta}{2} \propto \eta
\end{equation}
being directly proportional to $\eta$, that is, the fraction of detected photons.\\
\noindent To conclude these considerations, we extend the analysis of the micro-macro amplified
system in this highly attenuated scenario to the case where the injection of the single-photon in
the optical parametric amplifier occur with a non unitary efficiency $p < 1$. Such parameter 
represents the amount of matching (spectral, spatial, and temporal) between the optical mode of the amplifier
and the optical mode of the injected single-photon. To model this source of experimental imperfection,
the joint state between the two modes $\mathbf{k}_{A}$ and $\mathbf{k}_{B}$ before amplification is described by 
$\hat{\rho}^{-}_{p} = p \vert \psi^{-} \rangle_{AB} \langle \psi^{-} \vert + (1-p) \frac{\hat{I}_{A}}{2} \otimes \vert
0 \rangle_{B} \langle 0 \vert$, where $\hat{I}_{A} = \vert H \rangle_{A} \langle H \vert + \vert V \rangle_{A} \langle V \vert$
stands for a completely mixed polarization state and $\ket{0}_{B}\bra{0}$ represents the vacuum input state. 
By following the same procedure described for the $p=1$ case, the density
matrix of the joint micro-macro system after amplification and losses in the highly attenuated regime reads:

\begin{widetext}

\begin{equation}
\hat{\rho}_{\eta,p}^{AB} = \mathcal{N}_{\eta,p}^{-1} \left\{ \frac{2 p}{C^{2}} \frac{1}{1-t^{2}}
\begin{pmatrix} t^{2} & 0 & 0 & 0 \\ 0 & \frac{1}{2}\left( 1+t^{2}\right) & -\frac{1}{2}\left( 1+t^{2}\right) & 0\\
0 & -\frac{1}{2}\left( 1+t^{2}\right) & \frac{1}{2}\left( 1+t^{2}\right) & 0\\ 0 & 0 & 0 & t^{2} \end{pmatrix} +
(1-p) \Gamma \begin{pmatrix} t & 0 & 0 & 0 \\ 0 & t & 0 & 0 \\ 0 & 0 & t & 0 \\ 0 & 0 & 0 & t \end{pmatrix}
\right\}
\end{equation}

\end{widetext}

\noindent where $\mathcal{N}_{\eta,p}$ is the opportune normalization constant. In Fig.\ref{fig:concurrence_pmin1} (a)-(b) we show
the density matrix for a gain value $g=3$, for $\eta=10^{-4}$ and injection probabilities of $p=0.5$ and $p=0.25$. The effect of a decreasing
injection probability $p$ is the reduction of the off-diagonal terms and hence of the coherence terms. The application of
the Peres criterion on this density matrix gives a critical value of the injection probability $p_{crit} =
\frac{S^{2} (1-\eta)}{1 + S^{2} (1-\eta)}$. For $p>p_{crit}$, the micro-macro system in this
highly attenuated regime is entangled, while for $p \leq p_{crit}$ the system is separable. The same
result is confirmed by the calculation of the concurrence, which reads:
\begin{equation}
C(\hat{\rho}_{\eta,p}^{AB}) = \left\{ \begin{array}{ll} \frac{p (1-t^{2}) - (1-p) t S^2 (1-t^{2})}{
p(1 + 3 t^{2}) + 2 (1-p) t S^2 (1 - t^{2})} & \mathrm{for} \; p > p_{crit} \\
0 & \mathrm{for} \; p\leq p_{crit} \end{array}\right.
\end{equation}
In Fig.\ref{fig:concurrence_pmin1} (c) we report the plot of the concurrence as a function of the gain $g$ for several values
of the injection probability $p$ and $\eta=10^{-4}$. For decreasing $p$, the concurrence drops to $0$ for a lower value of the
gain. Furthermore, in Fig.\ref{fig:concurrence_pmin1} (d) we report the plot of the critical injection probability $p_{crit}$ as a
function of the gain $g$ and the transmission efficiency $\eta$. As the gain $g$ is increased, the value of the critical injection
probability increases up to a value close to $1$. This means that, for high values of the gain, an high injection efficiency is
requested to detect the entanglement with such measurement strategy.

\subsection{Spin-based entanglement criterion}
Another approach for a micro-macro entanglement test is based on the detection of the quantum Stokes operators, defined as:
$\hat{J}_{\vec{\pi}_{i}}^{B} = \hat{b}^{\dag}_{\vec{\pi}_{i}} \hat{b}_{\vec{\pi}_{i}} - \hat{b}^{\dag}_{\vec{\pi}^{\bot}_{i}} \hat{b}_{\vec{\pi}^{\bot}_{i}}$. For a micro-macro system, the following inequality, found by Simon et al. in Ref.\cite{Simo03}, holds for any separable state:
\begin{equation}
\vert \langle \vec{\hat{\sigma}}^{A} \cdot \vec{\hat{J}}^{B} \rangle \vert - \langle \hat{N}^{B} \rangle \leq 0
\end{equation}
where $\hat{N}^{B}$ is the photon number operator. For the micro-macro configuration under investigation, the following result \cite{Seka09,Seka10} holds:
\begin{equation}
\vert \langle \vec{\hat{\sigma}}^{A} \cdot \vec{\hat{J}}^{B} \rangle \vert - \langle \hat{N}^{B} \rangle = 2 \eta \geq 0
\end{equation}
thus violating the bound for separable states. Again, some entanglement survives for any value of the gain and of the losses parameter $\eta$, and the amount of entanglement is proportional to the number of detected photons. However such criterion is not feasible from an experimental point of view since the measurement of the Stokes operators requires perfect discrimination in the photon-number, as in the pseudo-Spin operator case.

\section{Conclusions}

In this paper we analyzed several classes of entanglement criteria for bipartite systems
of a large number of particles. In particular, we addressed a specific joint microscopic,
i.e. composed by a single particle, and macroscopic system based on optical parametric
amplification of an entangled photon pair. A first experimental entanglement test
on this system has been recently reported in Ref.\cite{DeMa08}. We analyzed in details
the conclusions that can be drawn on this experiment. The adopted entanglement criterion
in that paper allowed to infer the presence of entanglement after the amplification process
before losses in the detection apparatus. The validity of the test however requires
a specific assumption on the system that generates the micro-macro pair.
An a priori knowledge of the source is necessary in order to exclude a class of
separable states that can reproduce the obtained experimental results. One of the reason
for the necessity of this assumption is given by the exploited detection strategy, which presents
the feature of a POVM with an inconclusive outcome which depends on the measurement basis.
Such problem has been recently investigated in Ref.\cite{Vite10} within the context of a
nonlocality test in macroscopic systems. In that paper it was shown that a Bell test based
on a generalized dichotomic measurement, i.e. with an inconclusive outcome, allows to exclude
only a class of local hidden variables (LHV) models.

A more general approach to the micro-macro entanglement problem in the investigated system
is addressed in the rest of the paper. We discussed different entanglement criteria
which do not require any supplementary assumption on the source, and applied these
approaches to the micro-macro system based on optical parametric amplification. We first
derived a general bound for an entanglement criterion based on dichotomic operators. Then,
an approach based on deliberate attenuation of the multiphoton field to the single-photon regime,
already introduced in Ref.\cite{Eise04}, has been applied to our system.
This analysis allowed us to show that a fraction $\varepsilon$ of the original entanglement of the entangled
photon pair exists even in presence of losses, where $\varepsilon$ is proportional to the number of particles 
that survive the lossy process. As a further perspective, the system based on parametric amplification can lead to the
investigation of entanglement in a bipartite macroscopic macroscopic system \cite{DeMa10,Stob09}.

We acknowledge support by the ``Futuro in Ricerca'' Project HYTEQ,
and Progetto d'Ateneo of Sapienza Universit\`{a} di Roma.

\appendix

\section{Generalized micro-macro entanglement criterion for dichotomic operators}
\label{app:general_criterion}

In this appendix we demonstrate the inequality of Eq.(\ref{eq:general_criterion}), which gives
a generalized bound for an entanglement test in a micro-macro bipartite system and dichotomic
measurements. The proof is divided into two parts. First, a general treatement of dichotomic
measurements is applied to the derivation of an entanglement criterion with no auxiliary
assumption on the system under investigation. Then, the obtained results are applied to a
micro-macro scenario.

\subsection{General treatment of dichtomic measurements}
The density matrix of a separable state, composed by two
subsystems $A$ and $B$, can be written as:
\begin{equation}
\hat{\rho} =\sum_{i}p_{i}\left(\hat{\rho}_{i}^{A}\otimes\hat{\rho}_{i}^{B} \right)
\end{equation}
We restrict our attention to the set of dichotomic measurements, i.e. $(\pm 1)$ valued upon each
subsystem $\hat{O}_{A}$ and $\hat{O}_{B}$ respectively. The average value of
a generic measurement operator $\hat{O}^{j}= \hat{O}_{A}^{j}\otimes \hat{O}_{B}^{j}$
is given by $V^{j} = \mathrm{Tr} \left(\hat{\rho} \hat{O}^{j}\right)$,
where the superscript $j$ refers to a specific choice of the operator $\hat{O}^{j}$.
The average value of the $i-th$ component of the
decomposition of the density matrix reads:
\begin{equation}\label{eq:visibility_i}
v^{ij}=\mathrm{Tr}\Big(\left(\hat{\rho}_{i}^{A}\otimes\hat{\rho}_{i}^{B} \right)  \hat{O}^{j}\Big)
\end{equation}
The average value $V^{j}$ can then be reexpressed as:
\begin{equation}
V^{j}=\mathrm{Tr} \left(\sum_{i} p_{i} \left(\hat{\rho}_{i}^{A}\otimes \hat{\rho}_{i}^{B} \right)
\hat{O}^{j} \right) = \sum_{i} p_{i}\; v^{ij}
\end{equation}
The following inequality holds:
\begin{equation}
\left|V^{j}\right|=\left|\sum_{i} p_{i}\; v^{ij}\right|\leq
\sum_{i} p_{i}\; \left| v^{ij}\right|
\end{equation}
since $p_{i}\geq 0$. The sum of the average value over three different operators
$\hat{O}^{j}$, where $\left\{ j=1,\ldots,3 \right\}$, is given by the following expression:
\begin{equation}
\begin{aligned}
\sum_{j=1}^{3} \left\vert V^{j} \right\vert &\leq
\sum_{i} p_{i} \left| v^{i1}\right|+\sum_{i} p_{i}\left|
v^{i2}\right|+\sum_{i} p_{i}\left| v^{i3}\right| =\\
&= \sum_{i} p_{i} \Big(\left| v^{i1}\right|+\left| v^{i2}\right|
+\left| v^{i3}\right|\Big)
\end{aligned}
\end{equation}
By definition (\ref{eq:visibility_i}), we obtain for the $i-th$ component of the decomposition
for a separable state:
\begin{equation}
\begin{aligned}
v^{ij} &= \mathrm{Tr}\Big( \left(\hat{\rho}_{i}^{A}\otimes\hat{\rho}_{i}^{B}\right)
\left(\hat{O}^{j}_{A}\otimes \hat{O}^{j}_{B}\right) \Big) = \\
&=\mathrm{Tr}_{A} \left(\hat{\rho}_{i}^{A}
\hat{O}^{j}_{A}\right)\mathrm{Tr}_{B} \left(\hat{\rho}_{i}^{B} \hat{O}^{j}_{B}\right)= v^{ij}_{A}
\cdot v^{ij}_{B}
\end{aligned}
\end{equation}
Since $-1\leq v^{ij}_{B}\leq +1$ the following inequality holds:
\begin{equation}
\left|v^{ij}\right|=\left|v^{ij}_{A}\cdot
v^{ij}_{B}\right|\leq\left|v^{ij}_{A}\right|
\end{equation}
Hence for a generic separable state the following inequality
holds:
\begin{equation}
\label{eq:separable_inequality_intermediate}
\begin{aligned}
\sum_{j=1}^{3} \left\vert V^{j} \right\vert &\leq
\sum_{i} p_{i}
\Big(\left|v^{i1}\right|+\left|v^{i2}\right|+\left|v^{i3}\right|\Big) \\
&\leq
\sum_{i} p_{i}
\Big(\left|v^{i1}_{A}\right|+\left|v^{i2}_{A}\right|+\left|v^{i3}_{A}\right|\Big)
\end{aligned}
\end{equation}
where the $\left|v^{i1}_{A}\right|+\left|v^{i2}_{A}\right|+\left|v^{i3}_{A}\right|$ term is evaluated
over the density matrix $\hat{\rho}_{i}^{A}$ for subsystem $A$. The latter can be always
decomposed as:
\begin{equation}
\hat{\rho}_{i}^{A} = \sum_{n} q^{i}_{n} \vert \psi_{n} \rangle_{A} \, _{A}\langle \psi_{n} \vert
\end{equation}
where the set $\left\{q^{i}_{n} \right\}$ of probabilities satisfied the normalization condition
$\sum_{n} q_{n} = 1$. We can then derive the following inequality:
\begin{equation}
\begin{aligned}
\sum_{j=1}^{3} \left|v^{ij}_{A}\right| &=  \sum_{j=1}^{3} \left|
\mathrm{Tr} \left( \sum_{n} q^{i}_{n} \vert \psi_{n} \rangle_{A} \, _{A}\langle \psi_{n} \vert \hat{O}^{j}_{A}
\right) \right| = \\ &=\sum_{j=1}^{3}  \left|  \sum_{n} q^{i}_{n}
\mathrm{Tr} \left(\vert \psi_{n} \rangle_{A} \, _{A}\langle \psi_{n} \vert \hat{O}^{j}_{A}
\right) \right| \\ &\leq \sum_{j=1}^{3}  \sum_{n} q^{i}_{n} \left|
\mathrm{Tr} \left(\vert \psi_{n} \rangle_{A} \, _{A}\langle \psi_{n} \vert \hat{O}^{j}_{A}
\right) \right|
\end{aligned}
\end{equation}
Substituting this result in Eq.(\ref{eq:separable_inequality_intermediate}), by exploiting the normalization conditions  for the coefficients $\left\{ p_{i} \right\}$ and
$\left\{ q_{n}^{i} \right\}$  we obtain, due to convexity the following inequality for all bipartite separable states :
\begin{equation}
\label{eq:entanglement_criterium_separable}
\sum_{j=1}^{3} \left|V^{j}\right| \leq \max_{\vert \psi \rangle} \sum_{j=1}^{3}
\left| _{A}\langle \psi \vert \hat{O}^{j}_{A} \vert \psi \rangle_{A} \right|
\end{equation}
where the maximization is performed over all possible states of system $A$.

\subsection{Specific Micro-Macro case}
\label{sec:micro-macro}
We now specialize the result of previous section in the microscopic-macroscopic states,
i.e. when system $A$ is a single spin-$1/2$ particle. Let us make a specific choice for
the measurement operators $\left\{\hat{O}^{j}_{A} \right\}_{j=1}^{3}$. For a single spin-$1/2$ particle,
we choose the Pauli operators $\left\{ \hat{\sigma}^{j}_{A}\right\}_{j=1}^{3}$. Hereafter, we remove the
subscript $A$ in all the equations for simplicity of notation. The entanglement criterium
(\ref{eq:entanglement_criterium_separable})  for this choice of the system and operators
then reads:
\begin{equation}
\sum_{j=1}^{3} \left|V^{j}\right| \leq \max_{\vert \psi \rangle} \sum_{j=1}^{3}
\left| \langle \psi \vert \hat{\sigma}^{j} \vert \psi \rangle \right|
\end{equation}
We now need to maximize the righthand side of the latter equation over all possible
choices of single particle states $\vert \psi \rangle = \alpha \vert + \rangle + \beta \vert - \rangle$.
By applying the Lagrange multiplier method, the upper bound for the average of
the Pauli operators reads:
\begin{equation}
\sum_{j=1}^{3} \left| \mathrm{Tr} \left( \hat{\rho} \hat{\sigma}^{j} \right) \right| \leq \sqrt{3}
\end{equation}
Finally, we can write the following inequality for the joint microscopic-macroscopic system:
\begin{equation}
\sum_{j=1}^{3} \left| V^{j} \right| \leq \sqrt{3}
\end{equation}


\begin{thebibliography}{10}

\bibitem{Horo09}
R.~Horodecki, P.~Horodecki, M.~Horodecki, and K.~Horodecki,
\newblock Rev. Mod. Phys. {\bf 81}, 865 (2009).

\bibitem{Pere96}
A.~Peres,
\newblock Phys. Rev. Lett. {\bf 77}, 1413 (1996).

\bibitem{Horo96}
M.~Horodecki, P.~Horodecki, and R.~Horodecki,
\newblock Phys. Lett. A {\bf 223}, 1 (1996).

\bibitem{Duan00}
L.-M. Duan, G.~Giedke, J.~I. Cirac, and P.~Zoller,
\newblock Phys. Rev. Lett. {\bf 84}, 2722 (2000).

\bibitem{Brau05}
S.~Braunstein and P.~{van Loock},
\newblock Rev. Mod. Phys. {\bf 77}, 513 (2005).

\bibitem{Koro02}
N.~Korolkova, G.~Leuchs, R.~Loudon, T.~C. Ralph, and C.~Silberhorn,
\newblock Phys. Rev. A {\bf 65}, 053206 (2002).

\bibitem{Schn03}
R.~Schnabel {\em et~al.},
\newblock Phys. Rev. A {\bf 67}, 012316 (2003).

\bibitem{Koro05}
N.~Korolkova and R.~Loudon,
\newblock Phys. Rev. A {\bf 71}, 032343 (2005).

\bibitem{Simo03}
C.~Simon and D.~Bouwmeester,
\newblock Phys. Rev. Lett. {\bf 91}, 053601 (2003).

\bibitem{Chen02}
Z.-B. Chen, J.-W. Pan, G.~Hou, and Y.-D. Zhang,
\newblock Phys. Rev. Lett. {\bf 88}, 040406 (2002).

\bibitem{Juls01}
B.~Julsgaard, A.~Kozhekin, and E.~S. Polzik,
\newblock Nature {\bf 413}, 400 (2001).

\bibitem{Eise04}
H.~S. Eisenberg, G.~Khoury, G.~A. Durkin, C.~Simon, and D.~Bouwmeester,
\newblock Phys. Rev. Lett. {\bf 93}, 193901 (2004).

\bibitem{Cami06}
M.~Caminati, F.~{De Martini}, R.~Perris, F.~Sciarrino, and V.~Secondi,
\newblock Phys. Rev. A {\bf 73}, 032312 (2006).

\bibitem{Durk04}
G.~A. Durkin, C.~Simon, J.~Eisert, and D.~Bouwmeester,
\newblock Phys. Rev. A {\bf 70}, 062305 (2004).

\bibitem{DeMa05}
F.~{De Martini}, F.~Sciarrino, and V.~Secondi,
\newblock Phys. Rev. Lett. {\bf 95}, 240401 (2005).

\bibitem{DeMa08}
F.~{De Martini}, F.~Sciarrino, and C.~Vitelli,
\newblock Phys. Rev. Lett. {\bf 100}, 253601 (2008).

\bibitem{DeMa98}
F.~{De Martini},
\newblock Phys. Rev. Lett. {\bf 81}, 2842 (1998).

\bibitem{DeMa05a}
F.~{De Martini} and F.~Sciarrino,
\newblock Prog. Quant. Electr. {\bf 29}, 165 (205).

\bibitem{Scia05}
F.~Sciarrino and F.~{De Martini},
\newblock Phys. Rev. A {\bf 72}, 062313 (2005).

\bibitem{Scia07}
F.~Sciarrino and F.~{De Martini},
\newblock Phys. Rev. A {\bf 76}, 012330 (2007).

\bibitem{DeMa09}
F.~{De Martini}, F.~Sciarrino, and N.~Spagnolo,
\newblock Phys. Rev. Lett. {\bf 103}, 100501 (2009).

\bibitem{Durk04a}
G.~A. Durkin,
\newblock {\em Light and Spin entanglement},
\newblock PhD thesis, Corpus Christi College, University of Oxford, 2004.

\bibitem{Spag09}
N.~Spagnolo, C.~Vitelli, T.~{De Angelis}, F.~Sciarrino, and F.~{De Martini},
\newblock Phys. Rev. A {\bf 80}, 032318 (2009).

\bibitem{Seka09}
P.~Sekatski, N.~Brunner, C.~Branciard, N.~Gisin, and C.~Simon,
\newblock Phys. Rev. Lett. {\bf 103}, 113601 (2009).

\bibitem{Seka10}
P.~Sekatski, B.~Sanguinetti, E.~Pomarico, N.~Gisin, and C.~Simon,
\newblock arXiv:1005.5083 .

\bibitem{DeMa09a}
F.~{De Martini}, F.~Sciarrino, and N.~Spagnolo,
\newblock Phys. Rev. A {\bf 79}, 052305 (2009).

\bibitem{Vite10}
C.~Vitelli, N.~Spagnolo, L.~Toffoli, F.~Sciarrino, and F.~{De Martini},
\newblock Phys. Rev. A {\bf 81}, 032123 (2010).

\bibitem{DeMa10}
F.~{De Martini},
\newblock Found. Phys., DOI:10.1007/s10701-010-9417-3  (2010).

\bibitem{Stob09}
M.~Stobi\'{n}ska, {\em et al.},
\newblock arXiv:0909.1545v2  (2009).

\end{thebibliography}
\end{document}